\documentclass[a4paper,11pt]{article}

\usepackage{amssymb}
\usepackage{amsmath}
\usepackage{graphicx}
\usepackage{comment}

\begin{document}

\title{Electrical resistivity in warm dense plasmas beyond the average-atom model}

\author{J.C. Pain$^{\dag}$\footnote{phone: 00 33 1 69 26 41 85, email: jean-christophe.pain@cea.fr} and G. Dejonghe\footnote{CEA, DAM, DIF, B.P. 12, F-91297 Arpajon Cedex, France}}

\maketitle

%=======================
%		ABSTRACT
%=======================

\begin{abstract}
The exploration of atomic properties of strongly coupled partially degenerate plasmas, also referred to as warm dense matter, is important in astrophysics, since this thermodynamic regime is encountered for instance in Jovian planets' interior. One of the most important issues is the need for accurate equations of state and transport coefficients. The Ziman formula has been widely used for the computation of the static (DC) electrical resistivity. Usually, the calculations are based on the continuum wavefunctions computed in the temperature- and density-dependent self-consistent potential of a fictive atom, representing the average ionization state of the plasma (average-atom model). We present calculations of the electrical resistivity of a plasma based on the superconfiguration (SC) formalism. In this modeling, the contributions of all the electronic configurations are taken into account. It is possible to obtain all the situations between the two limiting cases: detailed configurations (a super-orbital is a single orbital) and detailed ions (all orbitals are gathered in the same super-orbital). The ingredients necessary for the calculation are computed in a self-consistent manner for each SC, using a density-functional description of the electrons. Electron exchange-correlation is handled in the local-density approximation. The momentum transfer cross-sections are calculated by using the phase shifts of the continuum electron wavefunctions computed, in the potential of each SC, by the Schr\"odinger equation with relativistic corrections (Pauli approximation). Comparisons with experimental data are also presented.
\end{abstract}

\maketitle

%======================
%	I. INTRODUCTION
%======================

\section{\label{sec1} Introduction}

The static electrical (DC) resistivity is important for the characterization of the plasma state. At high density, when the Spitzer theory becomes invalid, a consistent many-particle theory is required in order to evaluate the transport coefficients. The extended Ziman \cite{ZIMAN61} formulation of the electrical resistivity of liquid metals is based on linear response theory, where free electrons in a metal are uniformly accelerated until they collide with an ion and are scattered. In the $t-$matrix formulation of Evans \cite{EVANS73}, the resistivity reads \cite{PERROT87,PERROT99,ROZSNYAI08}:

\begin{equation}
\eta=-\frac{\hbar}{3\pi e^2Z^{*2}n_i}\int_0^{\infty}\frac{\partial f}{\partial\epsilon}(\epsilon,\mu)I(\epsilon)d\epsilon, 
\end{equation}

where $n_i=\rho\mathcal{N}/A$ is the matter density in cm$^{-3}$, $\rho$ the matter density in g.cm$^{-3}$, $A$ the atomic mass in g and $\mathcal{N}$ the Avogadro number. The Fermi-Dirac distribution and its derivative read respectively:

\begin{equation}
f(\epsilon,\mu)=\frac{1}{1+e^{\beta(\epsilon-\mu)}}\;\;\text{and}\;\;\frac{\partial f}{\partial\epsilon}(\epsilon,\mu)=-\beta f(\epsilon,\mu)[1-f(\epsilon,\mu)],
\end{equation}

where $\beta=1/(k_BT)$ and $\mu$ is the chemical potential. The function $I$ is defined by an integral:

\begin{equation}
I(\epsilon)=\int_{0}^{2k}S(q)\sigma(q)q^3dq,
\end{equation}

where $k^2=2m\epsilon/\hbar^2$, $m$ being the electron mass. The quantity $S(q)$ is the ionic structure factor and $\sigma(q)$ represents the scattering cross-section, $q^2=2k^2(1-\chi)$ and $\chi=\cos(\theta)$, where $\theta=(\vec{k},\vec{k'})$, $\vec{k}$ and $\vec{k'}$ being the momenta of the conduction electron respectively before and after the scattering event. Therefore, one has:

\begin{equation}\label{inti}
I(\epsilon)=2k^4\int_{-1}^1S(\sqrt{2k^2(1-\chi)})|a(k,\chi)|^2(1-\chi)d\chi,
\end{equation}

where $a(k,\chi)$, defined by $\sigma(q)=|a(k,\chi)|^2$, is the scattering amplitude. In fact, the resistivity takes now a Drude-like form

\begin{equation}
\eta=\eta_0\frac{1}{n_ea_0^3}\frac{\tau_0}{\tau}
\end{equation}

with $\eta_0=\hbar a_0/e^2$=21.74 $\mu\Omega$.cm, $a_0$=52.91772083 10$^{-10}$ cm being the Bohr radius. $n_e=Z^*n_i$ represents the electron density. The characteristic time $\tau_0=4\pi \epsilon_0\hbar a_0/e^2$ and the relaxation time $\tau$ are such that

\begin{equation}
\frac{\tau_0}{\tau}=\frac{a_0^2}{3\pi Z^*}<I>_{\epsilon}\;\;\text{where} \;\; <I>_{\epsilon}=-\int_0^{\infty}\frac{\partial f}{\partial\epsilon}(\epsilon,\mu)I(\epsilon)d\epsilon. 
\end{equation}

Note that any model which treats the continuum by a Thomas-Fermi approximation will not give the correct continuum density of states and not correctly incorporate the effects of electron-electron interactions in the electron-ion phase-shifts \cite{PAIN07}. Such a model is not reliable for the evaluation of the scattering cross-section. 

%=====================================
%	II. ELECTRONIC STRUCTURE MODEL
%=====================================

\section{\label{sec2} Electronic-structure model and electron-ion interaction}

The average-atom (AA) model we used relies on the following self-consistent calculation. Schr\"odinger's equation is solved in the Pauli approximation \cite{BLENSKI95,BOGDANOVICH07}, in which only first-order relativistic corrections are retained:

\begin{equation}\label{schr}
\left[-\frac{\hbar^2}{2m}\left[\frac{d^2}{dr^2}-\frac{l(l+1)}{r^2}\right]-eV_{\mathrm{scf}}+V_{\mathrm{mv}}+V_{\mathrm{D}}+V_{\mathrm{so}}\right]y_{\epsilon,l}(r)=\epsilon\; y_{\epsilon,l}(r),
\end{equation}

where $V_{\mathrm{mv}}$ is the mass-velocity term, $V_{\mathrm{D}}$ is the Darwin term, and $V_{\mathrm{so}}$ is the spin-orbit contribution. $y_{\epsilon,r}(r)$ is the radial part of the wavefunction multiplied by $r$. The potential splits into two parts $V_{\mathrm{scf}}(r)=V_{\mathrm{el}}(r)+V_{\mathrm{xc}}(r)$, where $V_{\mathrm{xc}}$ represents the exchange-correlation part, evaluated in the local-density approximation \cite{ICHIMARU87}, and $V_{\mathrm{el}}$ the electrostatic part. Outside the Wigner-Seitz (WS) sphere, \emph{i.e.} beyond the radius $r_{ws}$, a bound-electron wavefunction is given by:

\begin{equation}\label{jel}
y_{nl}(r)=A_{nl}\;r\;\kappa_l(-iKr), \;\; \text{with} \;\; K=\left[\frac{2m\epsilon}{\hbar^2}\left(1+\frac{\epsilon}{2E_0}\right)\right]^{1/2},
\end{equation}

where $\kappa_l$ is a modified spherical Bessel function of the third kind and $A_{nl}$ a constant fixed by the boundary condition of the wavefunction at $r_{ws}$ and the normalization. Such a boundary condition will be noted BC2 in Sec. \ref{sec4}, while the usual ``confining'' boundary condition $y_{nl}(r_{ws})=0$ will be referred to as BC1. Beyond $r_{ws}$, a free-electron radial wavefunction is written

\begin{equation}
y_{\epsilon,l}(r)=\left[\frac{2mK}{\pi\hbar^2}\left(1+\frac{\epsilon}{E_0}\right)\right]^{1/2}r\;\left[\cos[\delta_{l}(k)]\;j_l(Kr)-\sin[\delta_{l}(k)]\;n_l(Kr)\right],
\end{equation}

where $j_l$ and $n_l$ are the spherical Bessel functions of the first and second kinds respectively, $E_0$ is the rest-mass energy of the electron, and $\delta_{l}(k)$ is the phase shift, given by

\begin{equation}
\tan[\delta_l(k)]=\frac{k\;j_l'(kr_{ws})-\xi\;j_l(kr_{ws})}{k\;n_l'(kr_{ws})-\xi\;n_l(kr_{ws})},
\end{equation}

where $\xi$ is the logarithmic derivative of the radial wavefunction $y_{\epsilon,r}(r)/r$. The calculation of the phase-shifts $\delta_{l}(k)$ is painstaking and requires accurate numerical methods \cite{EBELING76,PAIN06a,PAIN06b}. One defines, for a fixed chosen $\Delta^{(1)}k$, the finite difference

\begin{equation}
\Delta^{(1)} \delta_k=\max_l\left\{\left|\delta_l(k)-\delta_l(k-\Delta^{(1)} k)\right|+\left|\delta_l(k+\Delta^{(1)} k)-\delta_l(k)\right|\right\}.
\end{equation}

Then, if $\Delta^{(1)} \delta_k\geq h$ ($h$ being an infinitesimal threshold), $N_k^{(1)}=\Delta^{(1)} \delta_k/h$ points are added in the interval $\left[k-\Delta^{(1)} k,k+\Delta^{(1)} k\right]$, which enables one to define

\begin{equation}
\Delta^{(2)} \delta_k=\left|\delta_l(k)-\delta_l(k-\Delta^{(2)} k)\right|+\left|\delta_l(k+\Delta^{(2)} k)-\delta_l(k)\right|
\end{equation}

where $\Delta^{(2)} k=2\Delta^{(1)} k/N_k^{(1)}$. Then, for each $k$ of the new grid, if $\Delta^{(2)} \delta_k\geq h$, then $N_k^{(3)}=\Delta^{(2)} \delta_k/h$ points are added in the interval $\left[k-\Delta^{(2)} k,k+\Delta^{(2)} k\right]$, etc. The iterative process goes on until the resonance is correctly described. The radial electron density is given by

\begin{equation}
n_{\mathrm{free}}(r)\approx \int_0^{\infty}d\epsilon f(\epsilon,\mu)\sum_{l=0}^{\infty}\frac{2(2l+1)}{4\pi}\frac{y_{\epsilon,l}^2(r)}{r^2}.
\end{equation}

The new potential $V_{\mathrm{el}}$, solution of Poisson's equation, is inserted in Eq. (\ref{schr}) and the process continues until convergence is reached.

%====================================
%	III. SCATTERING CROSS-SECTION
%====================================

\section{\label{sec3} Scattering cross-section}

The electron-ion cross-section for momentum transfer reads 

\begin{equation}\label{eics}
\sigma(q)=|a(k,\chi)|^2=\frac{1}{k^2}\left|\sum_{l=0}^{\infty}(2l+1)e^{i\delta_l(k)}\sin[\delta_l(k)]P_l(\chi)\right|^2,
\end{equation}

where $P_l(\chi)$ are the Legendre polynomials. It is useful to split the summation over $l$ in two parts in Eq. (\ref{eics}): the first summation running from 0 to $l_0$ (typically equal to 30) and the second one from ($l_0$+1) to $\infty$. The latter can be approximated by the classical expression of the cross-section of a screened charge (summed from 0 to $\infty$) minus the values for $l\leq l_0$. Assuming, at high electron energies, a classical scattering cross-section for a screening potential $\sigma(q)\approx A^2/(q^2+\lambda^2)^2$, this consists in writing \cite{PERROT87}

\begin{eqnarray}
\frac{1}{k}\sum_{l=l_0+1}^{\infty}(2l+1)e^{i\delta_l(k)}\sin[\delta_l(k)]P_l(\chi)&=&\frac{A}{2k^2}[\sum_{l=0}^{\infty}(2l+1)Q_l(\zeta)P_l(\chi)\nonumber\\
&&-\sum_{l=0}^{l_0}(2l+1)Q_l(\zeta)P_l(\chi)]\nonumber\\
&=&\frac{A}{2k^2}\left[\frac{1}{\zeta-\chi}-\sum_{l=0}^{l_0}(2l+1)Q_l(\zeta)P_l(\chi)\right], 
\end{eqnarray}

with $A=2mZ^*e^2/\hbar^2$ and $\zeta=1+\lambda^2/2k^2$, where $\lambda$ is a screening length. The quantity $Q_l(\zeta)$ represents Legendre function, evaluated using recursive relations for low $x$ and from an integral representation (see Ref. \cite{GRADSHTEYN80}, p. 1017, \S \; 8.821) for $x\geq$ 1.02. One gets

\begin{equation}
a(k,\chi)=\frac{1}{k}\sum_{l=0}^{l_0}(2l+1)P_l(\chi)\left[\sin[\delta_l(k)]e^{i\delta_l(k)}-\frac{A}{2k}Q_l(\zeta)\right]+\frac{A}{2k^2}\frac{1}{\zeta-\chi}.
\end{equation}

Since $e^{2 i\delta_l(k)}-1\approx 2i\delta_l(k)=i A Q_l(z)/k$ \cite{PERROT87}, and since $\delta_l(k)$ is known, for $l$=0, $\cdots$, $l_0$, one fits the ratio $\delta_{l_0}(k)/\delta_{l_0-1}(k)$ to $Q_{l_0}(z)/Q_{l_0-1}(z)$ in order to determine $z$ and then $\lambda^2=2k^2(z-1)$. In that way, $\lambda$ is a function of $k$. It can be numerically costly to solve this equation for the numerous values of $k$. In order to speed up the computation, one can use the asymptotic form $Q_n(z)\approx 2^n(n!)^2/[(2n)!(2n+1)z^{n+1}]$, which leads to $Q_{n+1}(z)/Q_n(z)\approx (n+1)/[(2n+3)z]$, \emph{i.e.}, for $n=30$, $z\approx 0.492063\;\delta_{29}/\delta_{30}$. The maximum energy of the continuum is taken to be $\epsilon_m=\mu+40\;k_BT$. Therefore, at high temperatures, it can be relevant to add the following correction to the resistivity:

\begin{equation}
\Delta \eta=\frac{1}{3\pi}\frac{\hbar}{e^2}\frac{n_i}{n_e^2}f(\epsilon_m,\mu)\int_0^{\sqrt{8\epsilon_m}}q^3S(q)\sigma(q)dq.
\end{equation}

\begin{table}
\begin{center}
\begin{tabular}{|c|c|c|c|c|} \hline \hline
$\rho/\rho_0$ & PY & DH & Rinker \cite{RINKER85} & BD \cite{BRETONNET88} \\\hline \hline
0.1 & 732.1 & 642.5 & 731.6 & 723.9 \\ \hline
0.2 & 234.4 & 158.0 & 233.9 & 224.3 \\ \hline
0.3 & 212.8 & 128.3 & 212.0 & 196.3 \\ \hline
0.4 & 180.8 & 101.1 & 179.9 & 163.4 \\ \hline
0.5 & 142.6 & 74.7 & 141.8 & 126.8 \\ \hline\hline
\end{tabular}
\end{center}
\caption{Resistivity $\eta$ ($\mu\Omega$.cm) of Al at T=5 eV for different structure factors: PY: Percus-Yevick (Hard Sphere), DH: Debye-H\"uckel, Rinker (Ref. \cite{RINKER85}, Eq. (15), p. 4211), and BD: Bretonnet-Derouiche \cite{BRETONNET88}, structure factor for the one component plasma (OCP).}\label{tab1}
\end{table}

\begin{table}
\begin{center}
\begin{tabular}{|c|c|c|c|c|c|c|c|c|c|c||c|} \hline \hline
$\rho/\rho_0$ & $T$ (eV) & $\eta_{ps}$ MS (\cite{PERROT99}) & $\eta_{ps}$ AA & $\eta_{BC1}$ AA & $\eta_{BC2}$ AA & $\eta$ SC \\
& & (\cite{PERROT99}) & (\cite{PERROT99}) & (This work) & (This work) & (This work)\\\hline \hline
0.1 & 5 & 810 & 405 & 718.0 & 731.6 & 642.6 \\ \hline
0.2 & 5 & 720 & 590 & 231.9 & 233.9 & 316.1 \\ \hline
0.3 & 5 & 490 & 318 & 205.9 & 212.0 & 237.4 \\ \hline
0.4 & 5 & 312 & 240 & 132.0 & 179.9 & 200.2 \\ \hline
0.5 & 5 & 203 & 165 & 106.1 & 141.8 & 152.4 \\ \hline
0.2 & 10 & 382 & 288 & 296.1 & 360.7 & 375.9 \\ \hline
0.3 & 10 & 296 & 240 & 198.8 & 200.5 & 208.7 \\ \hline
0.4 & 10 & 220 & 220 & 151.8 & 175.0 & 185.0 \\ \hline
0.5 & 10 & 164 & 164 & 129.6 & 149.2 & 158.5 \\ \hline\hline
\end{tabular}
\end{center}
\caption{Resistivity $\eta$ ($\mu\Omega$.cm) of Al: comparisons with Table II of Perrot and Dharma-Wardana \cite{PERROT99}. Their resistivities presented here use the pseudo-potential (ps) formulation. MS: Multiple Scattering and AA: Average Atom. BC1: the wavefunction cancels at $r_{ws}$, BC2: the wavefunction extends outside the WS sphere (jellium model, see Eq. (\ref{jel})). SC: Superconfiguration averaged.}\label{tab2}
\end{table}

\begin{table}
\begin{center}
\begin{tabular}{|c|c|c|} \hline \hline
 $T$=5 keV, $\rho$=1965 g/cm$^3$ & Resistivity $\eta$ ($\mu\Omega$.cm) \\ \hline \hline
Rinker \cite{RINKER85} A & 0.827 \\ \hline
Rinker \cite{RINKER85} B & 0.649 \\ \hline
This work (BC1) & 0.693 \\ \hline
This work (BC2) & 0.701 \\ \hline\hline
\end{tabular}
\end{center}
\caption{Comparisons with theoretical results of Rinker \cite{RINKER85} for Fe. Model A corresponds to a Thomas-Fermi-Dirac potential and the combined structure factor of the third column of Table \ref{tab1}. Model B uses a Thomas-Fermi-Dirac potential with a Debye-H\"uckel structure factor.}\label{tab3}
\end{table}

\begin{figure}
\begin{center}
\vspace{0.6cm}
\includegraphics[width=10cm]{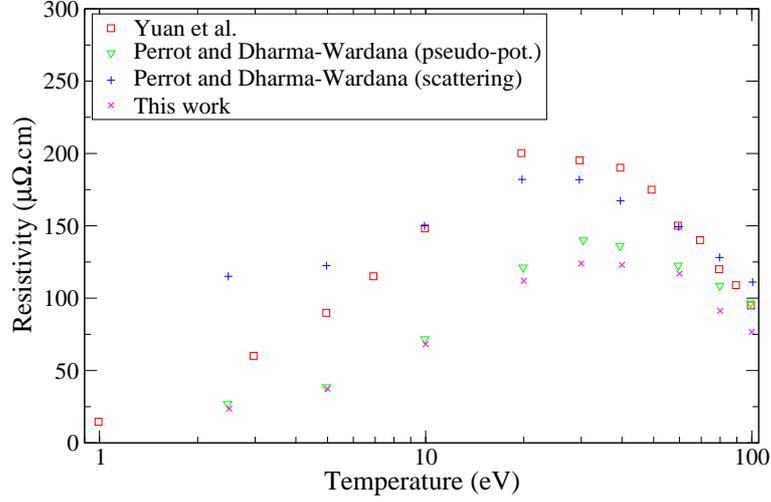}
\end{center}
\caption{Comparisons with calculations of Yuan \emph{et al.} \cite{YUAN96} and Perrot and Dharma-Wardana (Table I of Ref. \cite{PERROT99}: pseudo-potential + scattering formulations).}
\label{fig1}
\end{figure}

\begin{figure}
\begin{center}
\vspace{0.7cm}
\includegraphics[width=10cm]{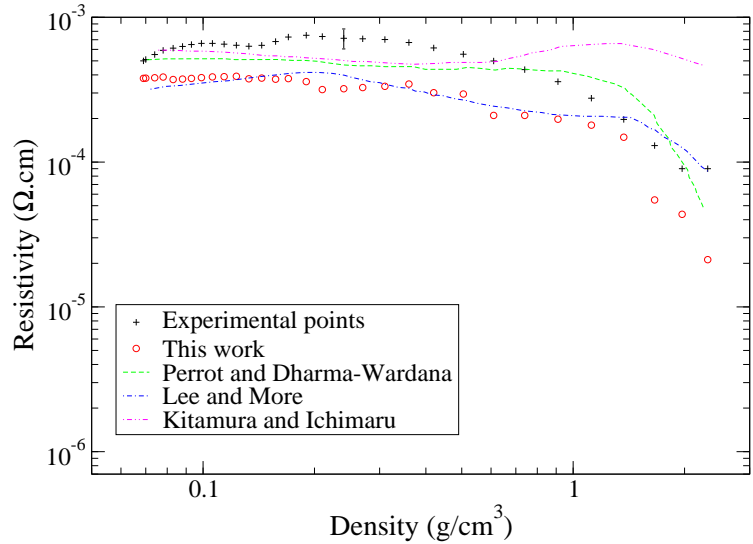}
\end{center}
\caption{Comparisons with the Benage experiment \cite{BENAGE99} for Al and with the calculations of Perrot and Dharma-Wardana \cite{PERROT87}, Lee and More \cite{LEE84} and Kitamura and Ichimaru \cite{KITAMURA95}.}
\label{fig2}
\end{figure}

%============================================================
%	IV. EFFECT OF BOUNDARY CONDITIONS AND STRUCTURE FACTOR
%============================================================

\section{\label{sec4} Comparisons with different theoretical models}

Rinker \cite{RINKER85} proposed a combined structure factor which has the virtue to approach Debye-H\"uckel limit as the core exclusion radius tends to 0, and the simple hard-sphere limit as the Debye radius tends to $\infty$. The structure factor is expected to be important at low temperature only ($T\le$ 10 eV according to \cite{PERROT99}), but as can be seen on Table \ref{tab1}, except for the Debye-H\"uckel case, the differences between Percus-Yevick, BD (structure factor proposed by Bretonnet and Derouiche \cite{BRETONNET88} for the OCP) and Rinker \cite{RINKER85} are quite small even at 5 eV. 

Table \ref{tab2} shows the impact of the value of the bound wavefunction at $r_{ws}$ (fifth and sixth columns). The maximum difference can reach 25 $\%$. In the case of iron at $T$=5 keV and $\rho$=1965 g/cm$^3$ (see Table \ref{tab3}), the difference between BC1 (the wavefunction cancels at $r_{ws}$) and BC2 (which allows a bound-electron wavefunction to extend outside the WS sphere, see Eq. (\ref{jel})) drops to 1.14 $\%$. Our results lie between the results of the models A and B of Rinker \cite{RINKER85}. In the following, we keep the condition BC2 (atom in a jellium). 

We can see in Table \ref{tab2} that our values decrease more rapidly with compression than the values of Ref. \cite{PERROT99} (with or without multiple scattering), and are almost always smaller. The comparisons of Fig. \ref{fig1} show that our results are closer to the pseudo-potential results including multiple scattering \cite{PERROT99} than to the scattering-type calculations \cite{PERROT99,YUAN96}. Some other approaches exist; for instance, Kuhlbrodt \emph{et al.} \cite{KUHLBRODT01} calculated the resistivity within the linear response method of Zubarev. This correlation method is applicable for arbitrary degeneracy of the plasma, \emph{i.e} from the low-temperature Spitzer regime up to the strongly coupled, degenerate domain where Ziman's approach is appropriate. It seems, however, that reliable resistivity calculations must include a unified description of the plasma composition and the density dependence of the inter-particle interactions \cite{KUHLBRODT01,RAMAZANOV07}.

%===============================================
%	V. COMPARISONS WITH EXPERIMENTAL VALUES
%===============================================

\section{\label{sec5} Comparisons with experimental values}

Resistivity measurements in plasma state are scarce so that the theoretical approach is in many cases the unique tool available in these studies. The measurement of resistivity along a thermodynamic path going from the melting point under normal conditions to a state corresponding to a compression 0.03 at T=26.5 eV in Al has been performed at Los Alamos by Benage \emph{et al.} \cite{BENAGE99} using the exploded wire method. Figure \ref{fig2} shows that a good agreement is obtained. The accuracy of the experimental results is generally within a factor of 2, which is of the same order of the typical discrepancy between theory and experiment. It is important to keep in mind that expression (\ref{inti}) neglects multiple-scattering processes, which was shown to lead to a difference of a factor 1.5 in the theoretical values \cite{PERROT99}. Our values are smaller than the results of Perrot and Dharma-Wardana \cite{PERROT87} and the results of the density-response model of Kitamura and Ichimaru \cite{KITAMURA95}. In the intermediate density range, our results agree fairly well with the analytical model of Lee and More \cite{LEE84}, which was rather unexpected.

%===========================================
%	VI. EFFECT OF POPULATION FLUCTUATIONS
%===========================================

\section{\label{sec6} Effect of population fluctations}

The number of relevant bound-electron configurations in a plasma can be immense, especially as the atomic number $Z$ increases. As a configuration is defined by orbitals occupied by an integer number of bound electrons, a superconfiguration (SC) is defined by super-orbitals, which are groups of orbitals close in energy (\emph{i.e.} whose energies differ by less than a small fraction of $k_BT$) and that are populated in all possible ways consistent with the Pauli exclusion principle. For instance $\Xi=(1s2s2p)^{10}(3s3p)^{7}(3d)^{2}(4s4p4d4f5s...)^{1}$ is a SC composed by four super-orbitals. A configuration is a particular SC in which each super-orbital contains only one sole orbital. A reasonable number, \emph{i.e.}, a few hundred, of SCs can contain a large number of configurations (especially for high values of $Z$). The value of thermodynamic quantity $X$ is given by

\begin{equation}
\langle X\rangle=\sum_{\Xi}W_{\Xi}X_{\Xi},
\end{equation}

where $W_{\Xi}$ is the probability of the configuration $\Xi$ and $\sum_{\Xi}W_{\Xi}=1$. A self-consistent calculation similar to the one described in Sec. \ref{sec2} is performed for each SC. Table \ref{tab2} displays the values of the resistivity averaged over the superconfigurations (third column).

A correction is required in the use of $Z^*$ instead of $Z_{\Xi}$ since this occurs as $Z_{\Xi}^2$ in the scattering cross-section. That is, we need to use $\langle Z_{\Xi}^2\rangle$ rather than $\langle Z_{\Xi}\rangle^2$. The fluctuations in $Z_{\Xi}$ are intimately related to the fluctuations of the electron and ion densities of the plasma. Assuming that ion fluctuations are too slow to follow the electron-density fluctuations which determine the electronic configurations and electron-scattering processes, we consider the fluctuations in the electron subsystem only. We have

\begin{equation}
<Z_{\Xi}^2>=<Z_{\Xi}>^2(1+n_i\chi_e/\beta),
\end{equation}

where the finite-temperature interacting electron-gas compressibility $\chi_e$ can be obtained from the electron-gas response function via the static structure factor since $\langle Z_{\Xi}\rangle n_i\chi_e=\beta S_{ee}(q=0)$.

Concerning the prefactor of the Ziman formula, since one has $\langle 1/Z_{\Xi}^2\rangle\geq\langle 1/Z_{\Xi}\rangle^2$ (Jensen's inequality for convex functions) and assuming that this dependence overcomes the one from the scattering cross-section, the resistivity averaged over SCs is expected to be larger than the one obtained from the AA model. However, it is important to keep in mind the fact that the latter prefactor should not be $1/\langle Z_{\Xi}^2\rangle$ but rather $\langle\frac{1}{Z_{\Xi}}\rangle\frac{1}{\langle Z_{\Xi}\rangle}$, which could impact the results. This is related to Rinker's remark \cite{RINKER88} that $1/Z^{*2}$ is in fact $1/(Z_0Z_i)$ where $Z_0$ can indeed be identified with $Z^*$ on variational grounds (Boltzmann equation), which is not the case of $Z_i$, number of charge carriers, involved in the relaxation time.
 
%======================
%	VII. CONCLUSION
%======================

\section{\label{sec7} Conclusion}

We presented calculations of the DC electrical resistivity of warm dense plasmas using the generalized Ziman formula. The calculations do not use any model potential, dielectric function or structure data, but proceed via first-principle calculations based on density-functional theory in the local-density approximation. The electronic structure of an average scatterer is determined solving Sch\"odinger's equation. The calculations require phase shifts which satisfy the finite-temperature Friedel sum rule. Different model structure factor have been tested, with two different boundary conditions of the wavefunctions. We discussed the approximations in the method arising from the use of an average-atom model and compared our results with other existing theoretical models. Our values are in fairly good agreement with experimental data. The method proposed here provides a self-contained and self-consistent approach to the electrical conductivity of plasmas of arbitrary degeneracy, structure and density. It enables one to perform calculations beyond the average-atom model, taking into account the fluctuations due to the electron configurations describing different ionization and excitation states.

\textbf{Acknowledgements}

The authors are grateful to J. Benage for providing his experimental data and to Ph. Arnault and G. Faussurier for helpful discussions.

\end{document}